\documentstyle[12pt]{article}

\newlength{\dinwidth}                                                           
\newlength{\dinmargin}                                                          
\catcode`@=11

\def\@citex[#1]#2{\if@filesw\immediate\write\@auxout{\string\citation{#2}}\fi
  \def\@citea{}\@cite{\@for\@citeb:=#2\do
    {\@citea\def\@citea{,\penalty\@m}\@ifundefined
      {b@\@citeb}{{\bf ?}\@warning
       {Citation `\@citeb' on page \thepage \space undefined}}%
\hbox{\csname b@\@citeb\endcsname}}}{#1}}

\def\citer{\@ifnextchar [{\@tempswatrue\@citexr}{\@tempswafalse\@citexr[]}}

\def\@citexr[#1]#2{\if@filesw\immediate\write\@auxout{\string\citation{#2}}\fi
  \def\@citea{}\@cite{\@for\@citeb:=#2\do
    {\@citea\def\@citea{--\penalty\@m}\@ifundefined
       {b@\@citeb}{{\bf ?}\@warning
       {Citation `\@citeb' on page \thepage \space undefined}}%
\hbox{\csname b@\@citeb\endcsname}}}{#1}}
\catcode`@=12

\def\bo{{\raise.15ex\hbox{\large$\Box$}}}               
\def\face{{\raise.2ex\hbox{$\displaystyle \bigodot$}\mskip-2.2mu \llap {$\ddot
        \smile$}}}                                      

\def\leftrightarrowfill{$\mathsurround=0pt \mathord\leftarrow \mkern-6mu
        \cleaders\hbox{$\mkern-2mu \mathord- \mkern-2mu$}\hfill
        \mkern-6mu \mathord\rightarrow$}       
\def\dvec#1{\vbox{\ialign{##\crcr
        \leftrightarrowfill\crcr\noalign{\kern-1pt\nointerlineskip}
        $\hfil\displaystyle{#1}\hfil$\crcr}}}           

\def\beq{\begin{equation}}
\def\eeq{\end{equation}}

\def\beqx{\begin{displaymath}}
\def\eeqx{\end{displaymath}}

\def\beql{\begin{eqnarray}}
\def\eeql{\end{eqnarray}}

\def\be{\begin{eqnarray}}
\def\ee{\end{eqnarray}}
\def\um{\frac{1}{2}}

\def\bee{\begin{eqnarray*}}
\def\eee{\end{eqnarray*}}
\def\nn{\nonumber}

\def\ss{\scriptstyle}
\def\sss{\scriptscriptstyle}
\def\mod{\mbox{mod}}

\begin{document}	

\begin{flushright}
NIKHEF/00-028\\
September 2000
\end{flushright}

\vspace{15mm}
\begin{center}
{\Large\bf\sc Open Descendants of $U(2N)$ Orbifolds at
rational radii}
\end{center}
\vspace{2cm}
\begin{center}
{\large A.N. Schellekens, N. Sousa}\\
\vspace{15mm}
{\em NIKHEF Theory Group\\
P.O. Box 41882, 1009 DB Amsterdam, The Netherlands} \\
\end{center}

\vspace{2cm}

\begin{abstract}

We construct explicitly the open descendants of some exceptional 
automorphism invariants of $U(2N)$ orbifolds. We focus on the case 
$N=p_1 \times p_2$, $p_1$ and $p_2$ prime, and on the automorphisms of the diagonal and charge conjugation invariants that exist for these values of $N$. These correspond to orbifolds of the circle with radius $R^2=2p_1/p_2$.
For each automorphism invariant we find two consistent Klein bottles, 
and for each Klein bottle we find a complete (and probably unique) set of boundary states. The two Klein bottles are in each case related to each other
by simple currents, but
surprisingly for the automorphism of the charge conjugation invariant neither
of the Klein bottle choices is the canonical (symmetric) one. 

\end{abstract}

\thispagestyle{empty}
\newpage \setcounter{page}{2}

\section{Introduction}

The method of open descendants \cite{sagnotti} enables one to construct an open string theory out of a closed one. Simply put, in this method one starts from the torus partition function of a modular invariant and left-right symmetric closed string theory. Then, since such theory has in general discrete symmetries, one can, by means of a Klein bottle projection, mod out a discrete symmetry group that contains at least the left-right symmetry (sometimes called world-sheet parity) as a subgroup. The resulting theory is however normally inconsistent due to massless tadpoles of unphysical fields. Consistency can then be restored by adequately adding open string sectors to the theory.

At first, and for some time, how to construct open descendants was known only for the case of the charge conjugation modular invariant, the so-called Cardy case. Gradually, attempts at extending the method to more general modular invariants began to spawn \cite{s&s}.  Recently, more progress has been achieved. In \cite{FOE}, universal formulas for open descendants of simple current invariants were given. Another class of solutions was studied in \cite{induct}, where a procedure for deriving the open descendants of modular invariants of the extension type was presented. 

These two developments cover a large number of cases but do not yet exhaust
all possibilities. Very little is known about the exceptional
invariants of automorphism type (apart from the $E_7$-invariant of $SU(2)$,
 level 16 studied in \cite{zuber}).
 In this letter we make a first attempt at solving this problem by explicitly constructing the open descendants of a theory with such exceptional invariants, namely orbifolds of $U(2N)$ conformal field theories. These theories have a large set of exceptional modular
invariants, related on the one hand to radius-reducing extensions of the
circle, and circles of rational radii $R^2=2p_1/p_2$ \cite{cloning}, and on the other hand to exceptional modular invariants of $SO(N)$, level 2 
\cite{galois} \cite{gannonBauto}. We will focus on two of these invariants here. 

Some work on boundary states for orbifold models was already done in \cite{affleck}. There the orbifold chiral algebra was studied in the continum limit and boundary coefficients for arbitrary radius were derived. This was done however in the absence of crosscaps. In this letter we consider the maximally extended chiral algebra at a rational point, along with the extra constraints imposed by crosscaps.

\section{Brief review of open descendants}

Before moving on to the $U(2N)$ orbifold theory we review some facts about open descendents and present key formulas.

The closed sector of a string theory is described by the 1-loop torus modular invariant partition function (MIPF), which is, dismissing integrations over the modular space and character arguments,
\be
T=\sum_{ij} \chi_i \;Z_{ij}\; \bar{\chi}_j.
\ee
$i,j$ label the chiral primary fields and $\chi_i,\bar{\chi}_j$ are their respective characters, the bar standing for anti-holomorphic. The matrix $Z_{ij}$ characterizes the various MIPF's of the conformal field theory (CFT). It is symmetric, has positive integer entries and commutes with the modular matrices $T$ and $S$ of the CFT.

The Klein bottle amplitude acts on this closed string spectrum as 
a projection.  
The Klein bottle can be seen in two different ways: direct and transverse channels. In the direct channel, it is viewed as a closed string propagating in a loop, coming back to itself with opposite orientation; this is interpreted as a partition function. In the transverse channel, the Klein bottle is viewed as a closed string propagating between two crosscap states; this is interpreted as a transition amplitude (and is denoted by a tilde). The two channels are related via modular transformations acting on the characters. This transformation is $S$ for the Klein bottle, which takes the form
\be
K=\sum_{i} K^i \chi_i \;\;\overrightarrow{{}_{\;\;\;\ss S\;\;\;}}\;\; 
\tilde{K}=\sum_m \Gamma_m^2 \chi_m, \label{DTKB}
\ee
with the requirement that $|K^i|=Z_{ii}$, which insures positivity and integrality of the unoriented closed sector partition function. Only fields that couple to their charge conjugate on the torus can propagate in the transverse channel \cite{scross}. These are called transverse fields and are labeled $m$. The crosscap coefficient $\Gamma_m$ is their 1-point function on the crosscap. Non-transverse fields always have $\Gamma=0$.

The open sector is described by the annulus and Moebius strip diagrams. In the direct channel, the annulus and Moebius strip have the interpretation of open strings propagating in a loop and coming back to itself, without or with an orientation flip respectively. They are the partition functions of the open sector. In the transverse channel, the interpretation is that of transition amplitudes for closed strings propagating between two boundary states and a boundary and crosscap state respectively. The channel transformations are $S$ for the annulus, and $P=\sqrt{T} S T^2 S \sqrt{T}$ for the Moebius strip. The expressions for the open sector are (transverse channels quantities marked with a tilde)
\be
&&A_{ab}=\sum_{i} A^i_{\;ab} \chi_i \;\;\overrightarrow{{}_{\;\;\;\ss S\;\;\;}}\;\;
\tilde{A}_{ab} = \sum_{m} B_{am} B_{bm} \chi_m, \label{DTA}\\
&&M_{a}= \pm\sum_{i} M_{\;a}^i \hat{\chi}_i  \;\;\overrightarrow{{}_{\;\;\;\ss P\;\;\;}}\;\; 
\tilde{M}_a = \pm\sum_{m} \Gamma_m B_{bm} \hat{\chi}_m,\label{DTMS}
\ee
where $a,b$ label the possible boundary conditions (types of D-branes). The annulus coefficients $A_{\;ab}^i$ are positive integers, as well as $M_{\;a}^i$, which are furthermore constrained by $M_{\;a}^i \leq A_{\;aa}^i, \; |M_{\;a}^i|=A_{\;aa}^i \;\mbox{mod}\; 2$, which insures positivity and integrality of the open sector partition function. The hatted characters in the Moebius denote the usual shifted argument.
$B_{am}$ are the boundary coefficients. They are the 1-point functions of the transverse fields $m$ on the disk with boundary condition $a$ and must satisfy the sewing constraints of \cite{sewing}. The open and closed sectors are further related via Chan-Paton factors and tadpole cancellation, but this is not relevant for our discussion and will not be exploited here.

Examining (\ref{DTKB}-\ref{DTMS}) we see that the relevant quantities on the direct and transverse channels are related by
\be
K^i=\sum_{m} S^i_{\;m} \Gamma_m^2,\;\; A^i_{\;ab}=\sum_m S^i_{\;m} B_{am} B_{bm},\;\; M^i_{\;a} = \sum_m P^i_{\;m} \Gamma_m B_{am}. \label{prob}
\ee
The problem boils down to solving these equations, under the restrictions of positivity and integrality, crosscap and sewing constraints. While positivity and integrality are straightforward to check, the crosscap and sewing constraints require knowledge of the fusing and brading matrices, which are not yet available for the orbifold theory. Nevertheless, the constraints imposed by positivity and integrality are in general very strong, and, as we shall see, are already enough to determine the crosscap and boundary coefficients in our cases.

For future use, we now define U-crosscap and reflection coefficients as
\be
U_m = {\Gamma_m}{\sqrt{S_{0m}}},\;\; R_{am}={B_{am}}{\sqrt{S_{0m}}}. \label{UR}
\ee

To find the solutions we start with an allowed Klein bottle and compute
the corresponding crosscap coefficients from it. Then we
look for non-negative integer
annuli $A^i_{\;aa}$, such that the corresponding boundary coefficients 
vanish
for fields that are not allowed in the transverse channel, and such that
a Moebius amplitude exists that is consistent with it. Finally we try
to assemble these boundary coefficients into a complete set, 
satisfying the condition $R_{am}R_{bm}^*=\delta_{ab}$ \cite{stanev},
and such that the off-diagonal annuli $A^i_{\;ab}$ are non-negative
integers. This can only be done in practice for small $N$, but 
it was possible to extrapolate the solution to arbitrary $N$ and prove
that it satisfies all relevant constraints in general. Furthermore
for small $N$ we found only one solution in each of the four cases
(two modular invariants, each with two Klein bottles). 
The search for diagonal annuli is however necessarily limited
to small integers, and therefore we cannot make 
rigorous claims regarding
the uniqueness of our solutions. 

\section{The orbifold theory}

$U(2N)$ conformal field theories describe a free boson compactified on a circle of radius $R^2=2N, \; N \in$ {\bf Z}. Using the procedure of \cite{orbifolds} we can build a new theory, the so-called orbifold theory, by modding the ${\mbox{\bf Z}}_2$ symmetry out of the $U(2N)$ CFT \cite{DVVV}.
The procedure can be inverted: by extending the orbifold theory with an apropriate simple current one gets back the original $U(2N)$ CFT.
The orbifold theory has exceptional MIPF's of the automorphism type when $N$ is both odd and a product of at least two different prime factors \cite{cloning}.

The fields of the orbifold theory are as follows:
\be
\begin{array}{ccc}
\mbox{Field} & \mbox{Weight} & \mbox{Description} \\ \hline
O & 0 & \mbox{vacuum} \\
J & 1 & \mbox{simple current} \\
\phi^l & {N}/{4} & \mbox{complex fields} \\
\phi_k & {k^2}/{4N} & \mbox{fixed points of $J$} \\
\sigma_l & {1}/{16} & \mbox{twisted sector} \\
\tau_l & {9}/{16} & \mbox{twisted sector} \\
\end{array}\nn
\ee
Label $l$ takes values 1,2 and label $k$ runs from 1 to $N-1$. The modular matrices $S$ and $P$ for this theory are presented in appendix \ref{SP}.

The fusion rules of the orbifold theory are invariant under the simultaneous action of an automorphism $\omega$ on all labels of the fusion coefficients: $N_{ij}^{\;\;k}=N_{\omega(i)
\omega(j)}^{\;\;\;\;\;\;\;\;\;\;\omega(k)}$. This automorphism acts non-trivially only on the $\phi_k$. Its action is described explicitly on appendix \ref{AR}. For our purpose, the important thing about it is that it can be used to write down an exceptional torus MIPF:
\be
T=\sum_{ij} \chi_i \; \delta_{i\omega(j)} \; \bar{\chi}_j. \label{torus}
\ee
We will focus on the case $N=p_1 \times p_2, \;p_1 < p_2$ and both prime. For this case it turns out that the fields\footnote{We denote by $\dot{p}$ multiples of $p$; zero included, when possible.} $\phi_{k, \; k=\dot{p_1},\dot{p_2}}$ self-couple on the torus, whereas the remaining $\phi_k$ couple amongst themselves crosswise. Since the $\phi_k$ are real, the transverse fields are then $O$, $J$ and $\phi_{k,\; k=\dot{p_1},\dot{p_2}}$, in total $p_1+p_2$ of them.

We can construct another exceptional MIPF replacing $\delta_{i\omega(j)}$ with the charge conjugation matrix $C_{i\omega(j)}$ in (\ref{torus}). We will call the two MIPF's ``diagonal + automorphism" (D+A) and  ``Cardy + automorphism" (C+A) respectively. In (C+A) $\phi^l, \sigma_l, \tau_l$ become transverse fields, raising the total number of these to $p_1+p_2+6$.

Having the torus MIPF, the next step is finding the Klein bottle projection. The Klein bottle has to be such that non-transverse fields do not propagate in the transverse channel; in other words, that their $\Gamma$ vanishes. Since $\Gamma$ can be immediately calculated by inverting (\ref{DTKB}), we conclude that the signs $\varepsilon_j=K^j/Z_{jj}$ of the direct channel Klein bottle have to be such that
\be
\Gamma^2_i  = \sum_j S_{ij} \varepsilon_j Z_{jj} = 0, \;\;\; \forall_i: Z_{ii^c} = 0. \label{sumrule}
\ee
In the (D+A) case this condition is satisfied for the
``trivial" Klein bottle projection, $K^i=1$ for all the fields coupling diagonally on the torus. In addition there is
a second Klein bottle with $K^i = -1$ on the four twist fields $\sigma_j$
and $\tau_j$ and $K^i=1$ for the other diagonal fields. There are several other Klein bottle choices that satisfy
the sum rule  (\ref{sumrule}), but if in addition we impose the Klein
bottle constraint of \cite{descendants} \cite{simpleklein} these are the
only two that are allowed. These two Klein bottle choices are related
by the action of the simple current $J$, by the mechanism explained  
in \cite{simpleklein}. 
In the (C+A) case we observe first of all that surprisingly the trivial choice
$K^i=1$ for all the fields coupling diagonally on the torus violates the sum rule (\ref{sumrule}). 
There are however several
Klein bottle choices satisfying the sum rule, and two of them also
satisfy the Klein bottle constraint. 
One has $K^{\phi_k}=-1$ when $k$ is an odd multiple of $p_1$ and $K^i=1$ for the remaining fields coupling diagonally on the torus, and the other has $K^{\phi_k}=-1$ when $k$ is an odd multiple of $p_2$ and $K^i=1$ for the remaining fields coupling diagonally on the torus. These two Klein bottle choices are again related by the action of a simple current, but
this time by $\phi^l$.

For the two aforementioned Klein bottles of (D+A) and (C+A) it was possible to construct a complete set of boundary states, which are now presented.

\section{D+A invariant}
\subsection{D+A U-crosscap and reflection coefficients}

For the (D+A) case and trivial Klein bottle, the U-crosscap coefficients are
\be
\begin{array}{c|cccccc}
 \; &  O & J & \phi^j & \phi_k & \sigma_j & \tau_j \\ \hline
U_i & \um\left( \frac{1}{\sqrt{p_1}}+\frac{1}{\sqrt{p_2}} \right) &
\um\left( \frac{1}{\sqrt{p_1}}-\frac{1}{\sqrt{p_2}} \right) &
0& \ldots, \underbrace{\frac{1}{\sqrt{p_2}}}_{k = 2\dot{p_1}}, \ldots,
\underbrace{\frac{1}{\sqrt{p_1}}}_{k = 2\dot{p_2}},\ldots 
& 0 & 0 \end{array}
\label{U}
\ee
The dots stand for zero entries. We defined the coefficients in such a way that for this Klein bottle they are all positive\footnote{We can always do this, since the problem is invariant under $U_m \rightarrow \varepsilon_m U_m,\;R_{am} \rightarrow \varepsilon_{m} R_{am}$.}. The Klein bottle with $-1$ for the twist fields has $U_O$ and $U_J$ interchanged and also a minus sign for $\phi_{k,k=2\dot{p_1}}$. Note that for both cases all non-transverse fields have vanishing U-crosscap coefficient, as expected. 

In order to determine the reflection coefficients, we first have to classify the possible boundary conditions. It turns out that three types of boundary conditions are possible:
\begin{itemize}

\item{Type $b$. This type contains two boundary conditions, $b_1$ and $b_2$, regardless of $N$.} 

\item{Type $a_1$. These split further into two subsets, $a_{1f}$ and $a_{1f}^\prime$ (with $f$ an odd integer ranging from $1$ to $p_1 - 2$). Each of these subsets contains thus $(p_1 - 1)/2$ boundary conditions, amounting to $p_1 - 1$ boundary conditions coming from this type of boundary.}

\item{Type $a_2$. Similar to type $a_1$. It splits into subsets $a_{2f}$ and $a^\prime_{2f}$ (with odd $f$ ranging from $1$ to $p_2 -2$ this time). It contains $p_2 - 1$ boundary conditions.}

\end{itemize}
There are $p_1 + p_2$ boundary conditions in total, as many as the transverse fields.

For both Klein bottle projections the reflection coefficients are
\be
\begin{array}{c|cccccc}
 \;&  O & J & \phi^l & \phi_k & \sigma_l & \tau_l \\ \hline
R_{b_1,i}  & \frac{1}{\sqrt{2p_1}} & \frac{-1}{\sqrt{2p_1}} & 0  & \ldots, \underbrace{\frac{2(-1)^n}{\sqrt{2p_1}}}_{k=2n{p_2}}, \ldots &  0 & 0   \\ \hline
R_{b_2,i}  & \frac{1}{\sqrt{2p_2}} & \frac{1}{\sqrt{2p_2}} & 0 & \ldots, \underbrace{\frac{2(-1)^n}{\sqrt{2p_2}}}_{k=2n{p_1}}, \ldots & 0 & 0 \\ \hline
R_{a_{1f},i} & \frac{1}{\sqrt{2p_1}} & \frac{-1}{\sqrt{2p_1}} & 0 & \ldots, \underbrace{{\frac{2 \cos{(\frac{\pi n f}{2p_1}})}{\sqrt{2p_1}}}}_{k=n {p_2}}, \ldots & 0 & 0   \\ \hline
R_{a^\prime_{1f},i} & \frac{1}{\sqrt{2p_1}} & \frac{-1}{\sqrt{2p_1}}& 0 & \ldots, \underbrace{{\frac{2(-1)^n  \cos{(\frac{\pi n f}{2p_1}})}{\sqrt{2p_1}}}}_{k=n {p_2}}, \ldots & 0 & 0   \\ \hline
R_{a_{2f},i} &\frac{1}{\sqrt{2p_2}} & \frac{1}{\sqrt{2p_2}} & 0 & \ldots, \underbrace{{\frac{2\cos{(\frac{\pi n f}{2p_2}})}{\sqrt{2p_2}}}}_{k=n {p_1}}, \ldots & 0 & 0   \\ \hline
R_{a^\prime_{2f},i} &\frac{1}{\sqrt{2p_2}} & \frac{1}{\sqrt{2p_2}} & 0 & \ldots, \underbrace{{\frac{2(-1)^n  \cos{(\frac{\pi n f}{2p_2}})}{\sqrt{2p_2}}}}_{k=n {p_1}}, \ldots & 0 & 0
\end{array} \label{R}
\ee
As expected, the non-transverse fields have vanishing reflection coefficients. Suppressing the zero columns in (\ref{R}), we come to an orthogonal square matrix. Orthogonality of $R$ is important, since it ensures that the completeness conditions of \cite{scross} are satisfied. Note that 
for the limiting case $f=p_l$ the 
boundaries $a_{lf}$ and $a'_{lf}$ are both equal to $b_l$. It is however
convenient to treat the $b$-boundaries separately. 


\subsection{D+A annulus and Moebius strip}

With the knowledge of the U-crosscap and reflection coefficients, we can now calculate the direct channel quantities, the annulus and Moebius strip coefficients. For briefness we present only the diagonal annulus $A^i_{\;aa}$, which contains the most important information. Using (\ref{DTA}) we get
\be
\begin{array}{c|ccccccc}
\; &  O & J & \phi^l & \phi_k & \sigma_l & \tau_l & \;\\ \hline
A^i_{\;b_1,b_1} & 1 & 1 & 1 & \ldots, \underbrace{2}_{k=\dot{p_1}},\ldots & 0 & 0 & \\
A^i_{\;b_2,b_2} & 1 & 1 & 1 & \ldots, \underbrace{2}_{k=\dot{p_2}},\ldots & 0 & 0 & \\

A^i_{\;a_{1f},a_{1f}} = A^i_{\;a_{1f}^{\prime},a_{1f}^{\prime}} & 1 & 1 & 0 & \ldots, \underbrace{1}_{k\pm f= {\ss even} \;\dot{p_1}}, \ldots, \underbrace{2}_{k= {\ss even} \;\dot{p_1}},\ldots & 0 & 0 &\\
A^i_{\;a_{2f},a_{2f}} = A^i_{\;a_{2f}^{\prime},a_{2f}^{\prime}} & 1 & 1 & 0 & \ldots, \underbrace{1}_{k\pm f= {\ss even} \;\dot{p_2}}, \ldots, \underbrace{2}_{k= {\ss even} \;\dot{p_2}},\ldots & 0 & 0 &\\
\end{array} \label{A}
\ee
Since the reflection coefficients are the same for the two Klein bottles, this is the annulus for both cases.
Evaluating (\ref{DTMS}) for the trivial Klein bottle projection gives us the Moebius strip
\be
\begin{array}{c|cccccccc}
\; & \;&  O & J & \phi^l & \phi_k & \sigma_l & \tau_l & \;\\ \hline
M^i_{\;b_1} & & 1 & 1 & 1 & \ldots, \underbrace{2}_{k={\ss odd}\;\dot{p_1}},\ldots & 0 & 0 & \\
M^i_{\;b_2} & & 1 & 1 & 1 & \ldots, \underbrace{2}_{k={\ss odd}\;\dot{p_2}},\ldots & 0 & 0 & \\
M^i_{\;a_{1f}} = M^i_{\;a_{1f}^\prime} & & 1 & 1 & 0 & \ldots, \underbrace{1}_{k \pm f = {\ss even}\;\dot{p_1}},\ldots & 0 & 0 & \\
M^i_{\;a_{2f}} = M^i_{\;a_{2f}^\prime} & & 1 & 1 & 0 & \ldots, \underbrace{1}_{k \pm f = {\ss even}\;\dot{p_2}},\ldots & 0 & 0 &
\end{array} \label{Mt}
\ee
For the Klein bottle with $-1$ for the twist fields the Moebius is
\be
\begin{array}{c|cccccccc}
\; & \;&  O & J & \phi^l & \phi_k & \sigma_l & \tau_l & \;\\ \hline
M^i_{\;b_1} & & -1 & -1 & 1 & \ldots, \underbrace{2}_{k={\ss odd}\;\dot{p_1}},\ldots & 0 & 0 & \\
M^i_{\;b_2} & & 1 & 1 & -1 & \ldots, \underbrace{-2}_{k={\ss odd}\;\dot{p_2}},\ldots & 0 & 0 & \\
M^i_{\;a_{1f}} = M^i_{\;a_{1f}^{\prime}} & & -1 & -1 & 0 & \ldots, \underbrace{1}_{k\pm f = {\ss even} \;\dot{p_1}},\ldots & 0 & 0 & \\
M^i_{\;a_{2f}} = M^i_{\;a_{2f}^{\prime}} & & 1 & 1 & 0 & \ldots, \underbrace{-1}_{k\pm f = {\ss even}\;\dot{p_2}},\ldots & 0 & 0 &
\end{array} \label{M}
\ee
Inspection shows that (\ref{A}-\ref{M}) respect the positivity and integrality conditions for the open sector. One can also show that (\ref{R}) leads to a positive integer off-diagonal annulus $A^i_{\;ab}$.

\section{C+A invariant}

\subsection{C+A U-crosscap and reflection coefficients}
We present the results for the Klein bottle with $K^{\phi_k}=-1$ when $k$ is an odd multiple of $p_1$ and $K^i=1$ for remaining the fields that couple diagonally on the torus. Since in the (C+A) case no quantity is proportional to the difference $p_1-p_2$, the results for the second Klein bottle are obtained by simply interchanging $p_1 \leftrightarrow p_2$ on the formulas below.

The U-crosscap coefficients are
\be
\begin{array}{c|cccccc}
\; &  O & J & \phi^l & \phi_k & \sigma_l & \tau_l \\ \hline
U_i & 
\frac{1}{2\sqrt{p_2}} & \frac{1}{2\sqrt{p_2}} & \frac{1}{2\sqrt{p_1}} & \ldots,\underbrace{\frac{1}{\sqrt{p_1}}}_{k=odd\;\dot{p_2}} ,\ldots,
\underbrace{ \frac{1}{\sqrt{p_2}}}_{k=even\;\dot{p_1}},\ldots & 0 & 0 \\
\label{Uc+a}
\end{array}
\ee
Next we classify the possible boundaries conditions. As in the (D+A) case we get a certain number fixed boundaries along with an expanding part which depends on the odd parameter $f$.
\begin{itemize}
\item{Type $a, b, c, d$. There are two $a$-type boundaries, $a$ and $a^\prime$. These are real (in the sense that their reflection coefficients turn out to be real). The boundaries $b,c$ and $d$ are unique but complex, thus admitting conjugate boundaries $b^\prime,c^\prime$ and $d^\prime$ respectively. In total these four types generate eight boundary conditions.}
\item{Type $E_{f}$. This type of boundary is real and splits into subtypes $E_{f}$ and $E_{f}^\prime$, with odd $f$ ranging from 1 to $p_1-2$, and contributes with a total of $p_1 - 1$ boundary conditions.}
\item{Type $F_{f}$. This type is complex and thus splits into $F_{f}$ and its conjugate $F_{f}^\prime$, with off $f$ ranging from 1 to $p_2-2$, and contributes with $p_2 - 1$ boundary conditions.}
\end{itemize}
The total number of boundaries is thus $p_1+p_2+6$, as many as the transverse fields, as expected.

The reflection coefficients are
\be 
\begin{array}{c|cccccc}
 \;&  O & J & \phi^j & \phi_k & \sigma_j & \tau_j \\ \hline
R_{a,i} & 
\frac{1}{\sqrt{8p_1}} & 
\frac{1}{\sqrt{8p_1}} & 
\frac{1}{\sqrt{8p_1}} & 
\ldots, \underbrace{\ss \frac{1}{\sqrt{2p_1}}}_{k=n{p_2}}, \ldots &  \frac{1}{\sqrt{8}} & 
\frac{1}{\sqrt{8}} \\ \hline
R_{a^\prime,i} & 
\frac{1}{\sqrt{8p_1}} & 
\frac{1}{\sqrt{8p_1}} & 
\frac{1}{\sqrt{8p_1}} &
\ldots, \underbrace{\ss \frac{1}{\sqrt{2p_1}}}_{k=n{p_2}}, \ldots &  \frac{-1}{\sqrt{8}} & 
\frac{-1}{\sqrt{8}} \\ \hline
R_{b,i} & 
\frac{1}{\sqrt{8p_1}} & 
\frac{1}{\sqrt{8p_1}} & 
\frac{-1}{\sqrt{8p_1}}  & 
\ldots, \underbrace{\ss \frac{(-1)^n}{\sqrt{2p_1}}}_{k=n{p_2}}, \ldots &  \frac{i\sigma_{1j}}{\sqrt{8}} & 
\frac{i\sigma_{1j}}{\sqrt{8}}  \\ \hline
R_{c,i} & 
\frac{1}{\sqrt{8p_2}} & 
\frac{-1}{\sqrt{8p_2}}& 
\frac{i\epsilon\sigma_{1j}}{\sqrt{8p_2}} & 
\ldots, \underbrace{\ss \frac{(-1)^n}{\sqrt{2p_2}}}_{k=2n{p_1}}, \ldots, \underbrace{\ss \frac{i}{\sqrt{2p_2}}}_{k=(2n-1){p_1}}, \ldots &  
\frac{e^{i\pi\sigma_{1j}/4}}{\sqrt{8}} & \frac{-e^{i\pi\sigma_{1j}/4}}{\sqrt{8}}  \\ \hline
R_{d,i} & 
\frac{1}{\sqrt{8p_2}} & 
\frac{-1}{\sqrt{8p_2}}& 
\frac{i\epsilon\sigma_{1j}}{\sqrt{8p_2}} &
\ldots, \underbrace{\ss \frac{(-1)^n}{\sqrt{2p_2}}}_{k=2n{p_1}}, \ldots, \underbrace{\ss \frac{i}{\sqrt{2p_2}}}_{k=(2n-1){p_1}}, \ldots&   
\frac{-e^{i\pi\sigma_{1j}/4}}{\sqrt{8}} & \frac{e^{i\pi\sigma_{1j}/4}}{\sqrt{8}} \\ \hline
R_{E_{f},i} & 
\frac{1}{\sqrt{2p_1}} & 
\frac{1}{\sqrt{2p_1}} & 
\frac{1}{\sqrt{2p_1}} &
., \underbrace{\ss \frac{ 2(-1)^n \cos{( \frac{\pi 2nf}{2p_1} )} } {\sqrt{2p_1}} }_{k=2n{p_2}}, 
., \underbrace{\ss  \frac{  2 \delta_{\!f} (-1)^n 
\sin{(\frac{\pi (2n-1)f}{2p_1})}} {\sqrt{2p_1}  }}_{k=(2n-1){p_2}},.& 
0 & 
0   \\ \hline
R_{E_{f}^\prime,i} & 
\frac{1}{\sqrt{2p_1}} & 
\frac{1}{\sqrt{2p_1}} & 
\frac{-1}{\sqrt{2p_1}}& 
., \underbrace{\ss  \frac{ 2(-1)^n \cos{( \frac{\pi 2nf}{2p_1} )} } {\sqrt{2p_1}} }_{k=2n{p_2}}, 
., \underbrace{\ss  \frac{  -2 \delta_{\!f} (-1)^n 
\sin{(\frac{\pi (2n-1)f}{2p_1})}} {\sqrt{2p_1}  }}_{k=(2n-1){p_2}},.& 
0 & 
0   \\ \hline
R_{F_{f},i} & 
\frac{1}{\sqrt{2p_2}} & 
\frac{-1}{\sqrt{2p_2}} & 
\frac{i\sigma_{1j}}{\sqrt{2p_2}} &
., \underbrace{\ss  \frac{ 2\cos{( \frac{\pi 2nf}{2p_2} )} } {\sqrt{2p_2}} }_{k=2n{p_1}},., \underbrace{\ss  \frac{  -2i \delta_{\!f} (-1)^n 
\sin{(\frac{\pi (2n-1)f}{2p_2})}} {\sqrt{2p_2}  }}_{k=(2n-1){p_1}},.& 
0 & 
0   \\ 
\end{array} \nn \ee \be
\epsilon=(-1)^{(N+1)/2},\;\;\; \delta_f=(-1)^{(p_1+f)/2},\;\;\; \sigma_{1j}=2\delta_{1j}-1. \label{Rc+a} 
\ee
Again, both the U-crosscap and reflection coefficients vanish on the non-transverse fields. The reflection coefficients for boundary conditions $b^\prime,c^\prime,d^\prime$ and $F_f^\prime$ are simply the complex conjugates of their unprimed counterparts.

\subsection{C+A annulus and Moebius strip}

\section{C+A annulus and Moebius strip}\label{AMc+a}
The diagonal annulus and Moebius strip that go with the U-crosscap and reflection coefficients of (\ref{Uc+a}-\ref{Rc+a}) are
\be
\begin{array}{c|cccccccccc}
\; &  O & J & \phi^1 & \phi^2 & \phi_k & \sigma & \sigma_2 & \tau_1 & \tau_2 & \;\\ \hline
A^i_{\;a,a} = A^i_{\;a^\prime,a^\prime}& 1 & 0 & 0 & 0 & \ldots, \underbrace{1}_{k=even\;\dot{p_1}},\ldots & 0 & 0 & 0 & 0 \\
A^i_{\;b,b} = A^i_{\;b^\prime,b^\prime}& 0 & 1 & 0 & 0 & \ldots, \underbrace{1}_{k=even\;\dot{p_1}},\ldots & 0 & 0 & 0 & 0 \\
A^i_{\;c,c} = A^i_{\;d,d}& 0 & 0 & 1 & 0 & \ldots, \underbrace{1}_{k=odd\;\dot{p_2}},\ldots& 0 & 0 &0 &0\\
A^i_{\;c^\prime,c^\prime} = A^i_{\;d^\prime,d^\prime} & 0& 0 & 0 &1&\ldots, \underbrace{1}_{k=odd\;\dot{p_2}},\ldots& 0&0&0&0\\
A^i_{\;E_{f},E_{f}} = A^i_{\;E_{f}^\prime,E_{f}^\prime}& 1 & 1 & 0 & 0 &  ., \underbrace{1}_{k\pm f=odd\;\dot{p_1}},., \underbrace{2}_{k=even\;\dot{p_1}},. & 0 & 0 & 0 & 0 \\
A^i_{\;F_{f},F_{f}} = A^i_{\;F_{f}^\prime,F_{f}^\prime}& 0 & 0 & 1 & 1 &  ., \underbrace{1}_{k\pm f=even\;\dot{p_2}},., \underbrace{2}_{k=odd\;\dot{p_2}},.& 0 & 0 & 0 & 0\\ \label{Ac+a}
\end{array}
\ee
\be
\begin{array}{c|cccccccccc}
\; &  O & J & \phi^1 & \phi^2 & \phi_k & \sigma & \sigma_2 & \tau_1 & \tau_2 & \;\\ \hline
M^i_{\;a} = M^i_{\;a^{\prime}}& 1 & 0 & 0 & 0 & \ldots, \underbrace{(-1)^{k/2}}_{k=even\;\dot{p_1}},\ldots & 0 & 0 & 0 & 0 \\
M^i_{\;b} = M^i_{\;b^{\prime}}& 0 & 1 & 0 & 0 & \ldots, \underbrace{-(-1)^{k/2}}_{k=even\;\dot{p_1}},\ldots & 0 & 0 & 0 & 0 \\
M^i_{\;c} = M^i_{\;d} & 0 & 0 & 1 & 0 & \ldots, \underbrace{1}_{k=odd\;\dot{p_2}},\ldots & 0 & 0 & 0 & 0 \\
M^i_{\;c^{\prime}} = M^i_{\;d^{\prime}} & 0 & 0 & 0 & 1 & \ldots, \underbrace{1}_{k=odd\;\dot{p_2}},\ldots & 0 & 0 & 0 & 0 \\
M^i_{\;E_{f}} & 1 & 1 & 0 & 0 & \ldots, \underbrace{(-1)^{k/2}}_{k \pm f=odd\;\dot{p_1}},\ldots & 0 & 0 & 0 & 0 \\
M^i_{\;E_{f}^{\prime}} & 1 & 1 & 0 & 0 & \ldots, \underbrace{-(-1)^{k/2}}_{k \pm f=odd\;\dot{p_1}},\ldots & 0 & 0 & 0 & 0 \\
M^i_{\;F_{f}} & 0 & 0 & \epsilon & -\epsilon & \ldots, \underbrace{1}_{k\pm f=even\;\dot{p_2}},\ldots & 0 & 0 & 0 & 0 \\
M^i_{\;F_{f}^{\prime}} & 0 & 0 & -\epsilon & \epsilon & \ldots, \underbrace{1}_{k\pm f=even\;\dot{p_2}},\ldots & 0 & 0 & 0 & 0 \\
\label{Mc+a}
\end{array}
\ee
Positivity and integrality of the diagonal annulus and Moebius strip is explicit. Again it can be shown that (\ref{Rc+a}) leads to positive integer off-diagonal annulus.

\section{Outlook}

Formulas (\ref{U}-\ref{Rc+a}) together with (\ref{Ac+a}-\ref{Mc+a}) summarize the information on the open descendants of the $N=p_1 \times p_2$ class of $U(2N)$ orbifold theories and they are the main result of this letter. The open descendants thus obtained satisfy the strong constraining requirements of positivity and integrality of the open and closed sector partitions functions, as well as the completeness condition
and the Klein bottle constraint. 
This provides strong evidence for the correctness of our {proposal}. It is also interesting to note that in the (C+A) case the natural Klein bottle choice $K^i=1$ for all the fields coupling diagonally on the torus is inconsistent.

A natural next step would be the extension of our results to more general 
automorphisms. It might also be worthwhile
try and understand these results from the point of view of the $U(2N)$ theory. In the $U(2N)$ theory the exceptional invariants are mapped into simple current invariants, whose properties are well under control. Relating the boundary and crosscap coefficients of the orbifold theory exceptional MIPF with those of the $U(2N)$ theory simple current MIPF would undoubtedly help understanding the construction of open descendants of exceptional MIPF's in more general cases. This is currently under study.


\begin{center}
{\bf Acknowledgements}
\end{center}

We would like to thank C. Schweigert, Ya.S. Stanev and L.R. Huiszoon for useful discussions. N.S. would like to thank Funda\c c\~ao para a Ci\^encia e Tecnologia for
financial support under the reference BD/13770/97. 

\vspace{10mm}

\appendix
\section{Automorphisms of the orbifold theory}\label{AR}

In this appendix we describe the fusion rule automorphisms present in the orbifold theory.

We denote the automorphism by $\omega$. It acts trivially on all fields other than $\phi_k$: $\;\omega(O;J;\phi^l;\sigma_l;\tau_l)= O;J;\phi^l;\sigma_l;\tau_l$. On the $\phi_k$ the action is as follows. Write $\phi_k$ as
\be
\phi_k = \left\{ \begin{array}{cc} N-2g& \mbox{for $k$ odd}
\\ 2g& \mbox{for $k$ even}
 \end{array} \right.
\ee
with $g$ taking the values $1,\ldots,(N-1)/2$. Now look for the smallest integer $m>1$ such that $m^2=1 \; \mod \; N$. For any $x$ define the (unique) number $[x]_N$, $0 \leq [x]_N \leq N/2$, such that $x \equiv \pm [x]_N \; \mod \; N$ for some choice of sign and define the permutation 
\be
\pi_m: \pi_m(g)  = [mg]_N \label{perm}.
\ee
The automorphism then acts on $\phi_k$ as
\be
\omega(\phi_{2g}) &= &\phi_{2[mg]_N} \nn \\
\omega(\phi_{N-2g}) &=&  \phi_{N-2[mg]_N}. 
\ee
The (D+A) torus is then
\be
T=\sum_g \chi_{\phi_{2g}} \bar{\chi}_{\phi_{2[mg]_N}} + \chi_{\phi_{N-2g}} \bar{\chi}_{\phi_{N-2[mg]_N}} +  \mbox{diagonal in the other fields}.
\label{Td+a}\ee
The (C+A) torus is obtained applying charge conjugation to (\ref{Td+a}).

\section{The modular matrices $S$ and $P$ of the orbifold theory for $N$ odd}\label{SP}
\be
S=\frac{1}{\sqrt{8N}}\; \times\;\;\;\;\;\;\;\;\;\;\;\;\;\;\;\;\;\;\;\;\;\;\;\;\;\;\;\;\;\;\;\;\;\;\;\;\;\;\;\;\;\;\;\;\;
\;\;\;\;\;\;\;\;\;\;\;\;\;\;\;\;\;\; \nn
\ee
\be
\begin{array}{c|cccccc}
\; & O & J & \phi^{l^\prime} & \phi_{k^\prime} & \sigma_{l^\prime} & \tau_{l^\prime}\\ \hline
O & 1 & 1 & 1 & 2 & {\ss \sqrt{N}} & {\ss \sqrt{N}} \\ 
J & 1 & 1 & 1 & 2 & -{\ss \sqrt{N}} & -{\ss \sqrt{N}} \\
\phi^l & 1 & 1 & -1 & 2(-1)^{k^\prime} & i\sigma_{ll^\prime} {\ss \sqrt{N}} & i\sigma_{ll^\prime} 
{\ss \sqrt{N}} \\ \phi_k & 2 & 2 & 2(-1)^k & 4 \cos{(2\pi\frac{kk^\prime}{2N})}& 0 & 0 \\
\sigma_l & {\ss \sqrt{N}} & -{\ss \sqrt{N}} & i\sigma_{ll^\prime}{\ss \sqrt{N}} & 0 & 
{\ss \sqrt{N}} e^{\frac{i\pi}{4}{\sigma_{ll^\prime} (-1)^{\sss \frac{N+1}{2}}}} & 
-{\ss \sqrt{N}} e^{\frac{i\pi}{4}{\sigma_{ll^\prime} (-1)^{\sss \frac{N+1}{2}}}} \\
\tau_l & 
{\ss \sqrt{N}} & -{\ss \sqrt{N}} & i\sigma_{ll^\prime}{\ss \sqrt{N}} & 0 & -{\ss \sqrt{N}}
e^{\frac{i\pi}{4}{\sigma_{ll^\prime} (-1)^{\sss \frac{N+1}{2}}}} & 
{\ss \sqrt{N}}e^{\frac{i\pi}{4}{\sigma_{ll^\prime} (-1)^{\sss \frac{N+1}{2}}}}
\end{array}\nn 
\ee We define $\sigma_{ll^\prime}=2\delta_{ll^\prime}-1$. This matrix is 
obtained from \cite{DVVV}, with a correction for the fixed point entries.
\be
P= \um \times\;\;\;\;\;\;\;\;\;\;\;\;\;\;\;\;\;\;\;\;\;
\;\;\;\;\;\;\;\;\;\;\;\;\;\;\;\;\;\;\;\;\;\;\;\;\;\;\;\;\;\;\;\;\;\;
\;\;\;\;\;\;\;\;\;\;\;\;\;\;\;\nn
\ee
\be
\begin{array}{c|cccccc}
\; & O & J & \phi^{l^\prime} & \phi_{k^\prime} & \sigma_{l^\prime} & \tau_{l^\prime}\\ \hline
O & 1 & 1 & \frac{1}{\sqrt{N}} & { \frac{1-(-1)^{k^\prime}}{\sqrt{N}}} & 0 & 0 \\ 
J & 1 & 1 & -\frac{1}{\sqrt{N}} & { -\frac{1-(-1)^{k^\prime}}{\sqrt{N}}}& 0 & 0 \\
\phi^{l} & \frac{1}{\sqrt{N}} &  -\frac{1}{\sqrt{N}}  & -\sigma_{ll^\prime}\; e^{i\pi \frac{N}{2}} & { \frac{(1+(-1)^{k^\prime})(-1)^{k^\prime /2}}{\sqrt{N}}} & 0 & 0 \\
\phi_k & {\frac{1-(-1)^{k}}{\sqrt{N}}}  & {-\frac{1-(-1)^{k}}{\sqrt{N}}}  &  {\frac{(1+(-1)^{k})(-1)^{k /2}}{\sqrt{N}}} & \frac{2(1-(-1)^{k+k^\prime})\cos{(2\pi\frac{kk^\prime}{4N})}}{\sqrt{N}}& 0 & 0 \\
\sigma_{l} & 0 & 0 & 0 & 0 & 
{\ss} e^{\frac{-i\pi}{8}{\sigma_{ll^\prime}}}  & 
{\ss} e^{\frac{3i\pi}{8}{\sigma_{ll^\prime}}} \\
\tau_{l} & 0& 0 & 0 & 0 & {} e^{\frac{3i\pi}{8}{\sigma_{ll^\prime}}} & {-} e^{\frac{-i\pi}{8}{\sigma_{ll^\prime}}}
\end{array} \nn
\ee

\end{document}